\documentclass[10pt]{article}
\begin{document}

\begin{center}
{\LARGE\bf Self interaction of spins in binary systems}
\vskip 1em
{\Large M\'{a}ty\'{a}s Vas\'{u}th$^{*}$ and Bal\'{a}zs Mik\'{o}czi$^{\dag}$}
\vskip 1.5em
{\small {}$^{*}$ \it Research Institute for Particle and Nuclear Physics, Budapest
114, P.O.Box 49, H-1525 Hungary}

{\small {}$^{\dag}$ \it Departments of Theoretical and Experimental Physics, University of
Szeged, Szeged 6720, Hungary}
\end{center}
\vskip 1em

\begin{abstract}
Beyond point mass effects various contributions add to the radiative
evolution of compact binaries. We present all the terms up to the 
second post-Newtonian order contributing to the rate of increase 
of gravitational wave frequency and the number of gravitational wave 
cycles left until the final coalescence for binary systems with spin, 
mass quadrupole and magnetic dipole moments, moving on circular orbit.
We evaluate these contributions for some famous or typical compact 
binaries and show that the terms representing the self interaction of 
individual spins, given for the first time here, are commensurable 
with the proper spin-spin contributions for the recently discovered 
double pulsar J0737-3039.
\end{abstract}

\section{Introduction}

Neutron star and black hole binary systems are among the significant sources 
of gravitational radiation which detection is expected by the gravitational 
wave observatories. The frequency of the radiation emitted
by these binaries is expected to be in the sensitivity range of the 
Earth-based interferometric detectors \cite{obs}. Recently observations 
are under way to directly detect such signals and analysis methods 
were specified for inspiral signals from binaries with 3-20 solar masses 
\cite{LIGOBH} and for setting upper limits on inspiral event rates for binary 
neutron stars using interferometer data \cite{LIGONS}. The detection of 
gravitational radiation from these compact binary systems is also expected by 
the Laser Interferometer Space Antenna \cite{LISA}. Alternative theories of 
gravity can be tested \cite{BBW} and parameters of spinning compact binaries 
can be estimated from these measurements.

The final coalescence of compact binaries is preceded by an inspiral phase
for which the post-Newtonian (PN) approximation provides a reliable 
description. This description is considered valid until the system reaches 
the innermost stable circular orbit. In \cite{BCV} the authors have discussed 
the problem of the failure of the PN expansion during the last stages of inspiral, 
called the intermediate binary black hole problem \cite{IBBH}, both for spinning 
and non-spinning black hole binaries on quasicircular orbits. Reliable results 
can be achieved by stopping the integration at the minimum of the energy as 
function of orbital frequency \cite{Blanchet02}. Tidal torques become important 
in latter stages of the inspiral \cite{Kochanek,BC}. 

The equations of motion were given to 3.5 PN order accuracy in the post-Newtonian 
regime with the inclusion of spin-orbit (SO) effects and their
first PN correction in \cite{Will}. Spin-spin (SS) \cite{Kidder},
quadrupole-monopole (QM) \cite{Poisson} and magnetic dipole - magnetic
dipole (DD) contributions \cite{IT} to the accelerations were also discussed.
The rate of the radiative change of otherwise conserved quantities, like the secular energy 
and angular momentum losses $\langle dE/dt\rangle$ and $\langle d\mathbf{J}/dt\rangle$,
characterizes the backreaction of gravitational waves escaping compact
binaries on the orbit, which were computed to leading order in \cite{Peters,PM} 
and its PN and 2PN corrections in \cite{DD} and \cite{GI}.

When the interaction of the spins $\mathbf{S}_{\mathbf{i}}$ with the orbit is taken 
into account, the total angular momentum $\mathbf{J}=\mathbf{L}+\mathbf{S}
_{\mathbf{1}}+\mathbf{S}_{\mathbf{2}}$ is still conserved, however the
orbital angular momentum $\mathbf{L}$ is not, due to spin precession.
Its magnitude $L$ remains constant as a consequence of the specific functional form of 
the precession equation \cite{GPV3}. The radiative change of $\langle dE/dt\rangle$ and 
$\langle dL/dt\rangle$ characterize the backreaction on the radial part of the motion 
which were computed in \cite{GPV3,RS}.

The SS contributions to $\langle dE/dt\rangle$ and $\langle dL/dt\rangle$
were derived in \cite{spinspin1,spinspin2}, where a description in terms
of the magnitude of an angular averaged orbital angular momentum $\bar{L}$
was introduced. This method was used in the discussion of the
QM \cite{quadrup} and DD contributions \cite{mdipole}. The
magnitude $L$ of the orbital angular momentum is not conserved due to the
spin precessions caused by spin-spin, quadrupole-monopole and magnetic
dipole - magnetic dipole interactions.

The SS contributions in $\langle dE/dt\rangle$ and $\langle dL/dt\rangle$ 
given in \cite{spinspin1} contained not only interaction terms between the two 
spins, but self-interaction spin (SS-self) terms. These are originated
from the terms proportional to 
$J_{SO}^{\left( 3\right) jl}\left( \mathbf{a}_{N}\right) J_{SO}^{\left(
3\right) jl}\left( \mathbf{a}_{N}\right) $ in $dE/dt$ and $\epsilon
^{ijk}J_{SO}^{\left( 2\right) jl}\left( \mathbf{a}_{N}\right) J_{SO}^{\left(
3\right) kl}\left( \mathbf{a}_{N}\right) $ in $d\mathbf{J}^{i}/dt$, where 
$J_{SO}^{\left( n\right) jl}$ denotes the $n^{th}$ derivative of the
spin-orbit contribution of the velocity quadrupole moment evaluated with the
Newtonian acceleration $\mathbf{a}_{N}$.

Here we compute the SS-self contributions to the rate of increase 
of the gravitational wave frequency $f$ and to the accumulated number of
gravitational wave cycles $\mathcal{N}$. For completeness we enlist all
other contributions to $df/dt$ and $\mathcal{N}$ to 2PN order, namely the
PN, SO, SS, QM, DD, 2PN and tail terms.

Due to the emission of gravitational waves the orbit of the binary tends to 
circularize \cite{Peters}. Accordingly we consider circular orbits for 
which the gravitational wave frequency is twice the orbital frequency 
\cite{Thorne}. In the next section we evaluate the rate of increase of $f$
which is given by the rate of change of the orbital angular frequency 
$\omega =\pi f$ under radiation reaction:
\begin{equation}
\left( \frac{d\omega }{dt}\right) ^{circ}=\left( \frac{dE}{d\omega }\right)
^{-1}\left\langle \frac{dE}{dt}\right\rangle ^{circ}\ ,  \label{omegadot}
\end{equation}
where the expression $dE/d\omega $ is found by differentiating 
$E=E\left( \omega \right)$. The various contributions to the secular energy 
loss 
\begin{eqnarray} \label{Eloss}
\left\langle \frac{dE}{dt}\right\rangle =\left\langle \frac{dE}{dt}
\right\rangle _{N}+\left\langle \frac{dE}{dt}\right\rangle
_{PN}+\left\langle \frac{dE}{dt}\right\rangle _{SO+tail}    
+\left\langle \frac{dE}{dt}\right\rangle
_{2PN+(SS-self)+S_{1}S_{2}+QM+DD}\  
\end{eqnarray}
were given in \cite{DD}-\cite{mdipole} and \cite{BlanchetSchafer}. 

Eq. (\ref{omegadot}) is immediately integrated since it is an ordinary 
differential equation in $\omega$. By a second integration we obtain the 
accumulated number of gravitational wave cycles $\mathcal{N}$. We evaluate 
the different contributions to $\mathcal{N}$ for some famous or typical 
compact binary systems and compare the magnitude of the SS-self term 
with other contributions. 

\section{Frequency evolution}

Introducing spherical coordinates the components of the acceleration 
${\bf a}$ in the system $({\bf\hat n},\hat{\lambda},{\bf\hat{L}_N})$ are
\begin{eqnarray} \label{acomp}
{\bf\hat n\cdot a} = \ddot{r}-r\omega^2 \ , \quad 
{\bf\hat\lambda\cdot a} = r\dot{\omega}+2\dot{r}\omega \ , \quad
{\bf\hat{L}_N\cdot a} = -r\omega\left(\hat\lambda\cdot \frac{d{\bf\hat{L}_N}}{dt}\right)\ , 
\end{eqnarray}
where ${\bf r}=r{\bf\hat n}$ is the separation vector, 
${\bf L_N}=\mu{\bf r\times\dot{r}}$ is the Newtonian orbital angular momentum, 
$\hat\lambda={\bf\hat{L}_N}\times {\bf\hat n}$ and a hat denotes a unit vector. 
The orbital angular velocity $\omega$ is introduced by the relation
${\bf v} = \dot r{\bf \hat n} + r\omega {\bf \hat \lambda}$.
For circular orbits $\dot{r}=\ddot{r}=0$ and $v^{3}=m\omega $ holds. Here 
$m=m_1+m_2$ is the total and $\mu=m_1m_2/m$ is the reduced mass of the system. 
Consequently $\left( m\omega \right) ^{2/3}$ is of PN order. The
radial projection of the acceleration yields the orbital angular velocity 
as $\mathbf{r\cdot a}=-r^{2}\omega ^{2}$ (\ref{acomp}). From the explicit 
form of the acceleration, with various contributions given in 
\cite{Kidder,Poisson,IT} we find $\omega =\omega \left(r\right)$ and then
$r=r\left( \omega \right)$,
\begin{eqnarray}\label{romega}
r(\omega ) &=&m(m\omega )^{-2/3}\Biggl\{1-\frac{3-\eta }{3}(m\omega )^{2/3} 
 -\frac{m\omega }{3}\sum_{i=1}^{2}\left( 2\frac{m_{i}^{2}}{m^{2}}+3\eta
\right) \frac{S_{i}}{m_{i}^{2}}\cos \kappa _{i}    
  \\
-&&\!\!\!\!\!\!\!\!\!\!\!\!\!\!(m\omega )^{4/3}\Bigg[-\eta \left( \frac{19}{4}+\frac{\eta }{9}\right) 
 +\frac{S_{1}S_{2}}{2\eta m^{4}}\left( \cos \gamma -3\cos \kappa _{1}\cos
\kappa _{2}\right)    
+\frac{1}{4}\sum_{i=1}^{2}p_{i}\left( 3\cos ^{2}\kappa _{i}-1\right) +
\frac{d_{1}d_{2}\mathcal{A}_{0}}{2\eta m^{4}}\Biggr]\Biggr\}\ . \nonumber
\end{eqnarray}
In Eq.(\ref{romega}) we have introduced the following notations.
The magnitude of the spins and magnetic dipole moments are denoted by $S_i$
and $d_i$. In a coordinate system with the axes 
$(\mathbf{\hat{c},\hat{L}\times \hat{c},\hat{L}})$, where $\mathbf{\hat{c}}$
is the unit vector in the $\mathbf{J\times L}$ direction the polar angles 
of the spins are $\kappa_i$ and $\psi_{i}$. 
In \cite{mdipole} other coordinate systems were introduced with the axes 
$(\mathbf{\hat{b}_{i},\hat{S}_{i}\times \hat{b}_{i},\hat{S}_{i}})$ 
with $\mathbf{\hat{b}_{i}}$ are unit vectors in the $\mathbf{S_{i}\times L}$ 
directions, respectively. In this system the polar angles of the the 
magnetic dipole moments $\mathbf{{d}_{i}\,}$ are $\alpha _{i}$ and $\beta _{i}$.
Moreover $\gamma =\cos ^{-1}(\mathbf{\hat{S}_{1}\cdot \hat{S}_{2})}$,
$\lambda =\cos ^{-1}(\mathbf{\hat{d}_{1}\cdot \hat{d}_{2})}$ and $\eta=\mu/m$.
The parameters $p_{i}=Q_{i}/m_{i}m^{2}$ characterize the quadrupolar 
contributions with the quadrupole-moment scalar $Q_{i}$ \cite{Poisson}.
In the last term $\mathcal{A}_{0}$ is defined as
\begin{eqnarray}\label{Bk}
\mathcal{A}_{0} &=&2\cos \lambda +3(\rho _{1}\sigma _{2}-\rho _{2}\sigma
_{1})\sin (\psi _{2}-\psi _{1})    
-3(\rho _{1}\rho _{2}+\sigma _{1}\sigma _{2})\cos (\psi _{2}-\psi _{1})\ , \\
&&\quad \rho _{i} =\sin \alpha _{i}\cos \beta _{i}\ , \qquad  
\sigma _{i} =\cos \alpha _{i}\sin \kappa _{i}+\sin \alpha _{i}\sin \beta
_{i}\cos \kappa _{i}\ . \nonumber
\end{eqnarray}
$E=E\left( \omega \right)$ is obtained by the combination of Eq. (\ref{romega}) 
with the expression of the velocity on circular orbits $v=r\omega $ in the energy 
integral $E=E\left( r,v\right)$:
\begin{eqnarray} \label{Eomega}
&&E(\omega ) = -{\frac{1}{2}}\mu (m\omega )^{2/3}\Biggl\{1-\frac{1}{4}\left(
3+\frac{\eta }{3}\right) (m\omega )^{2/3}    
+m\omega \sum_{i=1}^{2}\left( \frac{8}{3}\frac{m_{i}^{2}}{m^{2}}+2\eta
\right) \frac{S_{i}}{m_{i}^{2}}\cos \kappa _{i}    \\
&&\!\!\!\!\!\!\!\!\!\!\!\!\!\!+(m\omega )^{4/3}\Biggl[{\frac{1}{8}}(-27+19\eta -\frac{\eta ^{2}}{3}) 
 +\frac{S_{1}S_{2}}{\eta m^{4}}\left( \cos \gamma -3\cos \kappa _{1}\cos
\kappa _{2}\right)    
+\frac{1}{2}\,\sum_{i=1}^{2}p_{i}(3\cos ^{2}\kappa _{i}-1)+\frac{d_{1}d_{2}
\mathcal{A}_{0}}{\eta m^{4}}\Biggr]\Biggr\}\ .  \nonumber
\end{eqnarray}
The radiative evolution of the orbital angular frequency for circular orbits 
is deduced from Eqs. (\ref{omegadot}), (\ref{Eloss}) and (\ref{Eomega}) as 
\begin{eqnarray} \label{dotomega}
\left\langle \frac{d\omega }{dt}\right\rangle ^{circ} &=&\frac{96\eta
m^{5/3}\omega ^{11/3}}{5}\Biggl[1-\left( \frac{743}{336}+\frac{11}{4}\eta
\right) \left( m\omega \right) ^{2/3}  \\  
&+&\left( 4\pi -\beta \right) m\omega +\Biggl(\frac{34103}{18144}
+\frac{13661}{2016}\,\eta    
+\frac{59}{18}\,\eta ^{2}+\sigma \Biggr)\left( m\omega \right) ^{4/3}
\Biggr]\ ,  \nonumber
\end{eqnarray}
where
\begin{equation}
\sigma =\sigma _{S_{1}S_{2}}+\sigma _{SS-self}+\sigma _{QM}+\sigma _{DD}\ .
\end{equation}
Here $\beta $, $\sigma _{S_{1}S_{2}}$, $\sigma _{SS-self}$, $\sigma _{QM}$ 
and $\sigma _{DD}$ are the spin-orbit, spin-spin, self-interaction spin, 
quadrupole-monopole and magnetic dipole-dipole parameters:
\begin{eqnarray}
\beta &=&\frac{1}{12}\sum_{i=1}^{2}\frac{S_{i}}{m_{i}^{2}}\left( 113\frac{
m_{i}^{2}}{m^{2}}+75\eta \right) \cos \kappa _{i}\ ,  \label{beta} \\
\sigma _{S_{1}S_{2}}&=&\frac{S_{1}S_{2}}{48\eta m^{4}}(-247\cos \gamma
+721\cos \kappa _{1}\cos \kappa _{2})\ ,  \label{SS} \\
\sigma _{SS-self}&=&\frac{1}{96m^{2}}\sum_{i=1}^{2}\left( \frac{S_{i}}{m_{i}}
\right) ^{2}\left( 6+\sin ^{2}\kappa _{i}\right) \ ,  \label{self} \\
\sigma _{QM}&=&-\frac{5}{2}\sum_{i=1}^{2}p_{i}\left( 3\cos ^{2}\kappa
_{i}-1\right) \ ,  \label{QM} \\
\sigma _{DD}&=&-\frac{5}{\eta m^{4}}d_{1}d_{2}\mathcal{A}_{0}\ .  \label{siDD}
\end{eqnarray}
The N, PN, SO, SS, 2PN and tail contributions to Eq. (\ref{dotomega}) were 
verified to agree with the results of \cite{PoissonWill}, the QM and DD terms 
with \cite{Poisson} and \cite{IT}, respectively.

Eq. (\ref{dotomega}) is an ordinary differential equation in $\omega$.
To linear order in the perturbations all angular variables can be considered 
constants. Since all the angles appear in perturbative corrections they are given 
with sufficient accuracy to PN order, in which order they do not change \cite{GPV3}. 
Hence the time-evolution of $\omega$ for circular orbits can be obtained by 
an integration over time: 
\begin{eqnarray} \label{omega}
\omega (t) &=&\frac{\tau ^{-3/8}}{8m}\Biggl\{1+\left( \frac{743}{2688}+\frac{
11}{32}\eta \right) \tau ^{-1/4}+\frac{3}{10}\left( \frac{\beta }{4}-\pi \right) \tau ^{-3/8} \\
&&+\Biggl(\frac{1855099}{14450688}+\frac{56975}{258048}\eta   
+\frac{371}{2048}\eta ^{2}-\frac{3\sigma }{64}\Biggr)\tau ^{-1/2}\Biggr\}\ , 
\nonumber
\end{eqnarray}
where the dimensionless time variable $\tau =\eta (t_{c}-t)/5m$ is related
to the time $(t_{c}-t)$ left until the final coalescence. The accumulated orbital 
phase $\phi _{c}-\phi$ is given by a further integration:
\begin{equation}
\phi _{c}-\phi =\frac{5m}{\eta }\int \omega (\tau )d\tau \ .
\end{equation}
The accumulated number of gravitational wave cycles emerges as 
\begin{eqnarray}  \label{N}
\mathcal{N} &=&\frac{\phi _{c}-\phi }{\pi }=\frac{1}{\pi \eta }\Biggl\{\tau
^{5/8}+\left( \frac{3715}{8064}+\frac{55}{96}\eta \right) \tau ^{3/8}   
+\frac{3}{4}\left( \frac{\beta }{4}-\pi \right) \tau ^{1/4} \\
&&+\Biggl(\frac{9275495}{14450688}+\frac{284875}{258048}\eta   
+\frac{1855}{2048}\eta ^{2}-\frac{15\sigma }{64}\Biggr)\ \tau ^{1/8}\Biggr\}\ . 
\nonumber
\end{eqnarray}
The 2PN and tail contributions agree with those given in \cite{Blanchet95} and
the other terms with Eq. (4.16) of \cite{Kidder}.

We enlist all contributions to $\mathcal{N}$ in Table \ref{tab:a} in terms
of $\beta$ and $\sigma$, evaluated from Eq. (\ref{N}) for compact binaries. These 
are the double pulsar J0737-3039 \cite{pulsar1,pulsar2,pulsar3}, one neutron 
star - stellar mass black hole binary and two examples of galactic 
black hole binaries \cite{BBW}.

\begin{table}
\begin{center}
\begin{tabular}{lcccc}
\hline
& {J0737-3039} & & & \\
{PN Order} & $1.337M_{\odot }\ 1.25M_{\odot }$ 
  & {$1.4M_{\odot }\ 10M_{\odot }$} 
  & {$10^4M_{\odot }\ 10^5M_{\odot }$}
  & {$10^7M_{\odot }\ 10^7M_{\odot }$} \\
\hline
$f_{in}(Hz)$ & $10$ & $10$ & $4.199\times 10^{-4}$ & $1.073\times 10^{-5}$ \\
$f_{fin}(Hz)$ & $1000$ & $360$ & $3.997\times 10^{-2}$ & $2.199\times 10^{-4}$ \\
\hline
$N$ & $18310$ & $3580$ & $21058$ & $535$ \\
$PN$ & $475.8$ & $212$ & $677$ & $55$ \\ 
$SO$ & $17.5\beta $ & $14\beta $ & $36\beta $ & $4\beta $ \\
$SS-self,SS,QM,DD$ & $-2.1\sigma $ & $-3\sigma $ & $-5\sigma $ & $-\sigma $ \\
$Tail$ & $-208$ & $-180$ & $-450$ & $-48$ \\
$2PN$ & $9.8$ & $10$ & $18$ & $4$ \\ 
\hline
\end{tabular}
\caption{The accumulated number of gravitational wave cycles. The frequencies $f_{in}$ 
and $f_{fin}$ are given for the frequency domain of the detectors or up to the innermost 
stable circular orbit. The first two columns refer to the LIGO/VIRGO type detectors and 
the last two to the LISA bandwidth.}
\label{tab:a}
\end{center}
\end{table}

\section{Conclusions}

Various contributions add to the rate of increase of the gravitational wave 
frequency and the accumulated number of gravitational wave cycles.
Here we have presented the complete set of contributions to the evolution of 
gravitational wave frequency and to the accumulated number of gravitational 
wave cycles up to the 2PN order, namely the PN, SO, SS, QM, DD, tail and 2PN
terms, with the inclusion of the previously unknown self-interaction spin 
terms. These results add to the closed system of first order differential
equations governing the secular evolution of radiating compact binaries
already derived in \cite{GPV3,spinspin1,spinspin2,quadrup,mdipole} and
represent an important step towards a complete characterization of the
orbital evolution.

To comment on the importance of the self-interaction spin contributions 
we evaluate the spin parameters $\sigma _{S_{1}S_{2}}$ and $\sigma _{SS-self}$ 
for the double pulsar J0737-3039. The neutron stars in this double pulsar 
have average radii $15$ km, masses of $1.337 M_{\odot }$ and $1.25 M_{\odot }$, 
pulse periods of $22.7$ ms and $2773.5$ ms. By the Jenet-Ransom model 
\cite{JenetRansom} the angle $\kappa _{1}$ (in their notation $\lambda $) 
has two possible values: $\kappa _{1}=167^{\circ}\pm 10^{\circ }$ 
(Solution 1), and $\kappa _{1}=90^{\circ }\pm 10^{\circ }$ (Solution 2). 
The other angle $\kappa _{2}$ can be determined
by solving numerically the spin precession equations which was done for
black hole binaries in \cite{schnittman}. According to \cite{pulsar4,pulsar5} 
it is likely that wind-torques from the energetically dominant 
component have driven the spin axis of the other component to align with the 
direction of $\mathbf{L}$, causing $\kappa_{2}=0 $. The estimates for the
spin parameters are given in Table \ref{tab:b}. 

We conclude that the proper spin-spin contribution vanishes whenever one of 
the spins is aligned with the orbital angular momentum and the other spin is 
perpendicular to it. Thus $\sigma _{SS-self}$ is the only spin-spin contribution 
in this case. Moreover, though four orders of magnitude smaller than the
spin-orbit effects, already considered in \cite{O'Connell}, self-interaction
spin contributions are found to be comparable with the proper spin-spin
contributions. This is due to the fact that one of the spins is two orders of 
magnitude larger than the other.

The self-interaction spin contributions are more important when one of the 
spins is negligible compared to the other. The motion of a test particle orbiting
around a massive spinning body is described by the Lense-Thirring approximation.
Its first correction in the gravitational radiation is represented by the 
SS-self contribution of the higher spin.

\begin{table}
\begin{center}
\begin{tabular}{lrr}
\hline
{Spin parameter (order)} & {Solution 1} & {Solution 2} \\
\hline
$\beta \;$ & $-0.166$ & $0.001$ \\
$\sigma _{S_{1}S_{2}}\;(10^{-4})$ & $-0.372$ & $0$ \\
$\sigma _{SS-self}$\/$\;(10^{-4})$ & $0.298$ & $0.345$ \\
\hline
\end{tabular}
\caption{Spin parameters for the two solutions of the Jenet-Ransom model
representing the binary pulsar J0737-3039}
\label{tab:b}
\end{center}
\end{table}

\section*{Acknowledgments}
This work has been supported by OTKA grants T046939, F049429 and TS044665.
M.V. wishes to thank the organizers of the conference for their support.

\end{document}